\documentclass[12pt]{iopart}

\usepackage{graphicx}

\begin{document}

\title[Effective Singlet-Triplet Model]{Electronic Properties of the Effective Singlet-Triplet Model}

\author{Maxim M. Korshunov\dag\ \footnote[3]{mkor@iph.krasn.ru} and Sergei G. Ovchinnikov\dag \ddag}

\address{\dag\ L.V. Kirensky Institute of Physics, Siberian Branch of RAS, Krasnoyarsk,
660036, Russia}

\address{\ddag\ UNESCO Chair "New materials and technology",
Krasnoyarsk State Technical University, Krasnoyarsk, 660074,
Russia}

\begin{abstract}

In present work the effective singlet-triplet model for
$CuO_{2}$-layer in the framework of multiband p-d model of
strongly correlated electrons is obtained. The resulting
Hamiltonian has a form of generalized singlet-triplet t-t'-J
model for p-type superconductors and form of usual t-t'-J model
for n-type superconductors. In the mean field approximation in
X-operator representation we derived equations for Gorkov type
Green functions. The symmetry classification of the
superconducting order parameter in the case of tetragonal lattice
resulted in $d_{x^{2}-y^{2}}$- and $d_{xy}$-types of singlet
pairing for both p- and n-type superconductors while s-type
singlet pairing don't take place. Also normal paramagnetic phase
of effective singlet-triplet model was investigated and the
Fermi-type quasiparticle dispersion over Brillouin zone, density
of states and evolution of Fermi level with doping were obtained.

\end{abstract}



\maketitle

\section{Introduction}

Almost twenty years passed from the discovery of High-$T_{c}$
Superconductivity (HTSC) but for now there is no widely accepted
theory of pairing mechanism in cuprates of p- and n-type. The
necessary ingredient for discussing possible mechanisms of HTSC
is the band structure of the fermion-like quasiparticles.
However, it is a difficult subject for ab initio calculations due
to the strong electron correlations. For this reason we will use
a model approach and in order to get an agreement with
experimental data we will start with a realistic multiband p-d
model of transition metal oxides.

The Hubbard model is often used to study electronic structure of
strongly correlated electron systems (SCES). To take into account
the chemistry of metal oxides the Hubbard model is generalized to
the p-d model, a simplest version of such a model has been
proposed by Emery \cite{bib1} and Varma et al. \cite{bib2}. In
this 3-band p-d model only $d_{x^{2}-y^{2}}$ Cu and $p_{\sigma}$
O orbitals are considered. One of the important features omitted
in this model is asymmetry of n-type (electrons doped) and p-type
(holes doped) cuprates. Point is that the spin-exciton concerned
with singlet-triplet excitation of two-hole term occurs only in
p-type systems, but not in n-type \cite{bib3}. Other feature is a
non-zero occupancy of $d_{z^{2}}$ Cu orbitals \cite{bib4}. There
also a dependence between $T_{c}$ and occupancy of $d_{z^{2}}$
was found. Hence more realistic model of $CuO_{2}$-layer has to
include $d_{x^{2}-y^{2}}$- and $d_{z^{2}}$- orbitals on copper and
$p_{x}$-, $p_{y}$-, $p_{z}$- orbitals on oxygen as well. Such
model is the multiband p-d model that has been proposed by
Gaididei and Loktev \cite{bib5}:

\begin{equation}
H_{p-d} =\sum_{r}H_{d} (r) +\sum_{i}H_{p} (i) +\sum_{<r,
i>}H_{pd} (r,i) +\sum_{<i, j>}H_{pp}(i,j), \label{eq1}
\end{equation}
where
\begin{eqnarray}
\fl H_{d}(r) = \sum_{\lambda, \sigma}\left[
\varepsilon_{\lambda}^{d} \,d_{\lambda r \sigma}^{+} \,d_{\lambda
r \sigma} +\frac{U_{\lambda }^{d} }{2} n_{\lambda r}^{\sigma} n_{\lambda r}^{\bar{\sigma}} \right. \nonumber \\
\lo- \left. \sum_{\lambda', \sigma'}\left( J_{\lambda
\lambda'}^{dd} d_{\lambda r \sigma}^{+} d_{\lambda r \sigma'}
d_{\lambda' r \sigma'}^{+} d_{\lambda' r \sigma} - \sum_{r'}
V_{\lambda \lambda'}^{dd} n_{\lambda r}^{\sigma}
n_{\lambda' r'}^{\sigma'} \right) \right], \\
\fl H_{p}(i) = \sum_{\alpha, \sigma} \left[
\varepsilon_{\alpha}^{p} p_{\alpha i \sigma}^{+} p_{\alpha i
\sigma} + \frac{U_{\alpha }^{p} }{2} n_{\alpha i}^{\sigma}
n_{\alpha i}^{\bar{\sigma}} \right. \nonumber \\
\lo- \left. \sum_{\alpha',\sigma'}\left( J_{\alpha \alpha'}^{pp}
p_{\alpha i \sigma}^{+} p_{\alpha i \sigma'} p_{\alpha' i
\sigma'}^{+} p_{\alpha' i \sigma} - \sum_{i'}V_{\alpha
\alpha'}^{pp} n_{\alpha i}^{\sigma} n_{\alpha' i'}^{\sigma'} \right) \right], \\
\fl H_{pd}(r,i) = \sum_{\alpha, \lambda, \sigma, \sigma'} \left[
\left( t_{\lambda \alpha}^{pd} p_{\alpha i \sigma}^{+} d_{\lambda
r \sigma} + H.c. \right) + V_{\alpha \lambda}^{pd} n_{\alpha
i}^{\sigma} n_{\lambda r}^{\sigma'} \right], \\
\fl H_{pp}(i,j) = \sum_{\alpha, \beta, \sigma} \left[ t_{\alpha
\beta}^{pp} p_{\alpha i \sigma}^{+} p_{\beta j \sigma} + H.c
\right].
\end{eqnarray}
Here $r$ and $i$  are Cu and O sites, $\lambda={d_{x^{2}-y^{2}},
d_{z^{2}}}$ and $\alpha={p_{x}, p_{y}, p_{z}}$ are orbital
indexes on given copper and oxygen site respectively,
$\varepsilon^{d}$ and $\varepsilon^{p}$ are energies of
$d_{x^{2}-y^{2}}$- and $d_{z^{2}}$- holes on copper and $p_{x}$-,
$p_{y}$-, $p_{z}$- holes on oxygen. $U^{d}$, $U^{p}$ are
intra-atomic Coulomb interactions, $t^{pd}$ is hopping integral
for nearest neighbors copper-oxygen, $t^{pp}$ is hopping integral
for oxygen-oxygen, $V^{dd}$, $V^{pp}$, $V^{pd}$ are inter-atomic
Coulomb interactions and $J^{dd}$, $J^{pp}$ are exchange
integrals. Abbreviation "H.c." means Hermitian Conjugation and
"$<r, i>$" denotes that sum runs only over indices $r \neq i$.

Simplest calculations of this model have been made by exact
diagonalization of $CuO_{4}$ \cite{bib3} and $CuO_{6}$
\cite{bib6} clusters. It was shown that when $d_{z^{2}}$-orbitals
are neglected then the triplet with energy $\varepsilon_{2T}$
lies above singlet with energy $\varepsilon_{2S}$ on about 2 eV.
This lets us to neglect $d_{z^{2}}$-orbitals and we immediately
come back to the 3-band model. However, at approach of
$d_{z^{2}}$-orbital energy to $d_{x^{2}-y^{2}}$-orbital energy
the difference $\varepsilon_{2T}-\varepsilon_{2S}$ decreases and,
at the certain parameters, crossover of singlet and triplet take
place. Similar results  where obtained in \cite{bib7} by
self-consistent field method and in \cite{bib8} by perturbation
theory. All these facts give us cause for deeper investigation of
processes concerned with two-particle triplet.

\section{Formulation of Effective Model}

In this paper we will use Hubbard operators (or so-called
X-operators) $X_{f}^{p\,q} \equiv \left| p\right\rangle
\left\langle q\right| $ on site f instead of annihilation and
creation operators because they are a good tool in the case of
strong electron correlation. Also defining that $m\leftrightarrow
(p,q)$ numerate a quasiparticle described by the Hubbard operator
$X^{p\,q}$, and $\gamma_{\lambda \sigma} (m)$ is a parameter of
X-operator representation for the single-electron annihilation
operator with orbital $\lambda$ and spin $\sigma$ we can have the
next correspondence between annihilation (and creation operators)
and Hubbard operators:
\begin{equation}
a_{f \lambda \sigma} = \sum_{m} \gamma_{\lambda \sigma}(m)
X_{f}^{m}.
\end{equation}
Hermitian conjugation of Hubbard operator indicated by cross:
${X_{f}^{m}}^{+}$.

\begin{figure}[h]
\begin{center}
\includegraphics[width=0.7\textwidth]{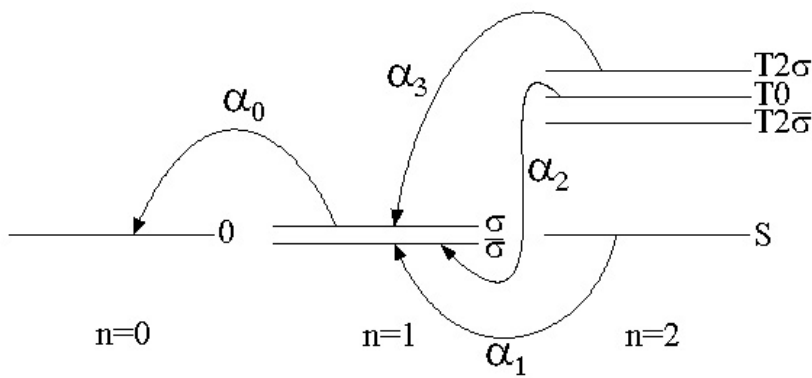}
\end{center}
\caption{\label{fig1}Quasiparticle excitations between state in
full singlet-triplet model (excitations for only one spin
projections are shown).}
\end{figure}

To calculate quasiparticle band structure in SCES the Generalized
Tight-Binding (GTB) method has been proposed \cite{bib9.1}. This
method combines the exact diagonalization of a cell part of the
Hamiltonian and the perturbation treatment of the intercell part
in the X-operator representation. In paper \cite{bib9.2} the
consequent development of GTB method for $La_{2}CuO_{4}$ with a
$CuO_{6}$ cluster as elementary cell was given. The problem of
non-orthogonality of nearest clusters' molecular orbitals was
solved in direct way by constructing explicitly Vanier functions
on $d_{x^{2}-y^{2}}, d_{3z^{2}-r^{2}}, p_{x}, p_{y}, p_{z}$ -
five orbitals' initial basis of atomic states. In a new symmetric
basis one-cell part of Hamiltonian are factorized allowing to
symmety classification of all possible effective one-particle
excitations in $CuO_{2}$ plane as transitions from n-th hole term
to (n+1)-hole term. The X-operators are constructed in the
Hilbert space that consists of a vacuum $n_{h}=0$ state  $| 0
\rangle$, single-hole $| \sigma \rangle = \left\{ | \uparrow
\rangle, | \downarrow \rangle \right\}$ molecular orbital of
$b_{1g}$ symmetry, two-hole singlet $| S \rangle \equiv |
\uparrow, \downarrow \rangle$ of $^{1}A_{1g}$ symmetry and
two-hole triplet $| TM \rangle$ (where $M=+1,0,-1$) of
$^{3}B_{1g}$ symmetry states. In the X-operator basis the
multiband p-d model Hamiltonian is given by:
\begin{eqnarray}
\fl H = \sum_{f} \left( \varepsilon_{1} \sum_{\sigma} X_{f}^{\sigma \sigma} + \varepsilon_{2S}X_{f}^{S S} + \varepsilon_{2T} \sum_{M} X_{f}^{TM TM} \right) \nonumber \\
\lo + \sum_{<f, g>, \sigma}\left[ t_{fg}^{0 0} X_{f}^{\sigma  0}
X_{g}^{0 \sigma} + 2 \sigma t_{fg}^{0 b} \left(
X_{f}^{\sigma 0} X_{g}^{\bar{\sigma} S} + X_{f}^{S \bar{\sigma}} X_{g}^{0 \sigma} \right) + t_{fg}^{b b} X_{f}^{S \bar{\sigma}} X_{g}^{\bar{\sigma} S} \right] \nonumber \\
\lo + \sum_{<f, g>, \sigma}\ t_{fg}^{a a} \left( \sigma \sqrt{2} X_{f}^{T0 \ \bar{\sigma}} - X_{f}^{T2\sigma \ \sigma} \right) \ \left( \sigma \sqrt{2} X_{g}^{\bar{\sigma} \ T0} - X_{g}^{\sigma \ T2\sigma} \right) \nonumber \\
\lo + \sum_{<f,\ g>,\ \sigma}\ t_{fg}^{a b} \left[ \left( \sigma
\sqrt{2} X_{f}^{T0 \ \bar{\sigma}} - X_{f}^{T2\sigma \ \sigma}
\right) \ \left( -\upsilon X_{g}^{0 \sigma} + 2\sigma \gamma_{b}
X_{g}^{\bar{\sigma} S} \right) + H.c. \right] \label{eq2}
\end{eqnarray}
Here the energies $\varepsilon_{1}$, $\varepsilon_{2S}$ and
$\varepsilon_{2T}$ are counted off from chemical potential $\mu$,
and $t_{fg}$ indexes $0$, $a$ and $b$ relevant to quasiparticle
in lower singlet ($0$), in higher singlet ($b$) and in higher
triplet ($a$) Hubbard's bands. The scheme of the levels and
available quasiparticle excitations between them are presented in
figure \ref{fig1}.

The t-t'-J model is derived by exclusion the intersubband hopping
between low (LHB) and upper (UHB) Hubbard subbands for the
Hubbard model \cite{bib10} and for 3-band p-d model \cite{bib11}.
We write the Hamiltonian in the form
\begin{equation}
H=H_{0}+H_{1},
\end{equation}
where the excitations via the charge transfer gap are included in
$H_{1}$. Then we define an operator
\begin{equation}
H\left(\epsilon \right)=H_{0}+\epsilon H_{1}
\end{equation}
and make the unitary transformation
\begin{equation}
\tilde{H} \left( \epsilon \right) =\exp{\left(-\rmi \epsilon
\hat{S} \right)} H\left( \epsilon \right) \,\exp{\left( \rmi
\epsilon \hat{S} \right)}.
\end{equation}

Vanishing linear in $\epsilon$ component of $ \tilde{H} \left(
\epsilon \right) $ gives the equation for matrix $\hat{S}$:

\begin{equation}
H_{1} + \rmi \left[ H_{0} ,\hat{S} \right]=0. \label{eq3}
\end{equation}

The effective Hamiltonian is obtained in second order in
$\epsilon$ and at $\epsilon=1$ is given by:
\begin{equation}
\tilde{H}=H_{0}+\frac{1}{2} \rmi \left[ H_{1}, \hat{S} \right].
\label{eq4}
\end{equation}

It is convenient to express the matrix $\hat{S}$ in terms of
X-operators \cite{bib12}. Thus, for the multiband p-d model
(\ref{eq1}) with electron doping (n-type systems) we obtain the
usual t-J model:
\begin{equation}
\fl H_{t-J}=\sum_{i, \sigma}\varepsilon_{1} X_{i}^{\sigma \sigma}
+ \sum_{<i, j>, \sigma} t_{ij}^{0 0} X_{i}^{\sigma 0} X_{j}^{0
\sigma} + \frac{1}{2} \sum_{<i, j>} J_{ij} \left( \vec{S}_{i}
\vec{S}_{j} - \frac{1}{4} n_{i} n_{j} \right),
\end{equation}
where spin operators $\vec{S}_{f}$ and number of particles
operators $n_{f}$ can be expressed in terms of Hubbard operators
as follows:
\begin{equation}
\vec{S}_{i} \vec{S}_{j} - \frac{1}{4} n_{i} n_{j} = \frac{1}{2}
\sum_{\sigma}\left( X_{i}^{\sigma \bar{\sigma}}
X_{j}^{\bar{\sigma} \sigma} - X_{i}^{\sigma \sigma}
X_{j}^{\bar{\sigma} \bar{\sigma}} \right).
\end{equation}

For p-type systems effective Hamiltonian has the form of a
singlet-triplet t-t'-J model \cite{bib12}:
\begin{equation}
\tilde{H}=H_{0}+H_{t}+H_{J}, \label{eq5}
\end{equation}
where $H_{0}$ (unperturbated part of the Hamiltonian), $H_{t}$
(kinetic part of $\tilde{H}$) and $H_{J}$ (exchange part of
$\tilde{H}$) are given by the following expressions:

\begin{eqnarray}
\fl H_{0} = \sum_{i}\left( \varepsilon_{1} \sum_{\sigma} X_{i}^{\sigma \sigma} + \varepsilon_{2S} X_{i}^{S S} + \varepsilon_{2T} \sum_{M} X_{i}^{TM TM} \right), \\
\fl H_{t} = \sum_{<i, j>, \sigma} t_{ij}^{b b} X_{i}^{S \bar{\sigma}} X_{j}^{\bar{\sigma} S} + \sum_{<i, j>, \sigma} t_{fg}^{a a} \left( \sigma \sqrt{2} X_{i}^{T0 \ \bar{\sigma}} - X_{i}^{T2\sigma \ \sigma} \right) \left( \sigma \sqrt{2} X_{j}^{\bar{\sigma} \ T0} - X_{j}^{\sigma \ T2\sigma} \right) \nonumber \\
\lo + \sum_{<i, j>, \sigma} t_{ij}^{a b} 2 \sigma \gamma_{b} \left[ X_{i}^{S \bar{\sigma}} \left( \sigma \sqrt{2} X_{j}^{\bar{\sigma} \ T0} - X_{j}^{\sigma \ T2\sigma} \right) +H.c. \right], \\
\fl H_{J} = \frac{1}{2} \sum_{<i, j>} \left( J_{ij} + \delta
J_{ij} \right) \left( \vec{S}_{i} \vec{S}_{j} - \frac{1}{4} n_{i}
n_{j} \right) - \frac{1}{2} \sum_{<i, j>, \sigma} \delta J_{ij}
X_{i}^{\sigma \sigma} X_{j}^{\sigma \sigma}.
\end{eqnarray}

The $J_{ij}$ is the exchange integral:
\begin{equation}
J_{ij} = 4 \frac{ \left( t_{ij}^{0 b} \right)^{2}}{E_{ct}},
\label{eq6}
\end{equation}
and $\delta J_{ij}$ is the correction to it (dependent on the
triplet's contribution):
\begin{equation}
\delta J_{ij} = 2 \upsilon^{2} \frac{\left(t_{ij}^{a b}
\right)^{2}}{E_{ct}}. \label{eq7}
\end{equation}
For the nearest neighbors ($i=0$, $j=1$) the estimation gives:
\begin{equation}
\frac{\delta J_{ij}}{J_{ij}} \sim 10^{-2}.
\end{equation}
Here $E_{ct}$ is the charge-transfer energy gap (similar to $U$
in the Hubbard model, $E_{ct} \approx 2 eV$ for cuprates).

Previously the motion of triplet holes and the simplest version
of singlet-triplet model have been studied in \cite{bib13,bib14}.
But these authors used the 3-band p-d model \cite{bib1,bib2}
where the difference between energy of singlet and energy of
triplet $\varepsilon_T-\varepsilon_S$ is about 2 eV. This fact
leads to negligible contribution of singlet-triplet excitations in
low-energy physics. But in multiband p-d model
$\varepsilon_T-\varepsilon_S$ depends strongly on various model
parameters, particularly on distance of apical oxygen
($p_z$-orbitals) from planar oxygen ($p_x$- and $p_y$-orbitals),
energy of apical oxygen, difference between energy of
$d_{z^2}$-orbitals and $d_{x^2-y^2}$-orbitals
($\varepsilon_{d_{z^2}}-\varepsilon_{d_{x^2-y^2}}$). For the
realistic values of model parameters
$\varepsilon_T-\varepsilon_S$ appears to be less or equal to 0.5
eV (see \cite{bib3} and \cite{bib9.2}). It is this small value of
singlet-triplet splitting that let us believe that
singlet-triplet excitations has a non-negligible contribution to
band structure and also can give new superconducting pairing
channel.

As easily can be noted the resulting Hamiltonian (\ref{eq5}) is
the generalization of t-J model to account of two-particle
triplet state. But the inclusion of this triplet leads to such
significant changes in Hamiltonian as renormalization of exchange
integral (\ref{eq7}) and appearing of $X_{i}^{\sigma \sigma}
X_{j}^{\sigma \sigma}$ term in $H_{J}$.

More significant feature of effective singlet-triplet model
(\ref{eq5}) is the asymmetry for n- and p-type systems. This fact
is known experimentally. In particular, the holes suppress
antiferromagnetism more strongly than electrons. It was observed
in $La_{2-x}Sr_{x}CuO_{4}$ in compare with $Nd_{2-x}Ce_{x}CuO_{4}$
\cite{bib15}. For n-type systems the usual t-J model takes place
while for p-type superconductors with complicated structure on
the top of the valence band the singlet-triplet transitions plays
an important role. In first case we have spin-fluctuation pairing
mechanism (see review \cite{bib16}). In the second case in
addition to spin-fluctuations described by $H_{J}$ we also have
pairing mechanism due to singlet-triplet excitations. Really,
let's look at structure of $H_{t}$. There we can see terms like
$X_{i}^{S\,\bar{\sigma}} X_{j}^{\bar{\sigma} \,T0} $ which can be
identically written according to multiplication rule of Hubbard
operators:
\begin{equation}
X_{i}^{S \bar{\sigma}} X_{j}^{\bar{\sigma} T0} = X_{i}^{S
\bar{\sigma}} X_{j}^{\bar{\sigma} S} X_{j}^{S T0}.
\end{equation}

Here we can see three processes: creation ($X_{i}^{S
\bar{\sigma}}$) and annihilation ($X_{j}^{\bar{\sigma} S}$) of
the hole at different sites i and j, and spin-exciton $X_{j}^{S
T0}$. This spin-exciton can play a role of intermediate boson in
superconducting pairing.

\section{Green Functions for Effective Singlet-Triplet Model}

\begin{figure}[h]
\begin{center}
\includegraphics[width=0.6\textwidth]{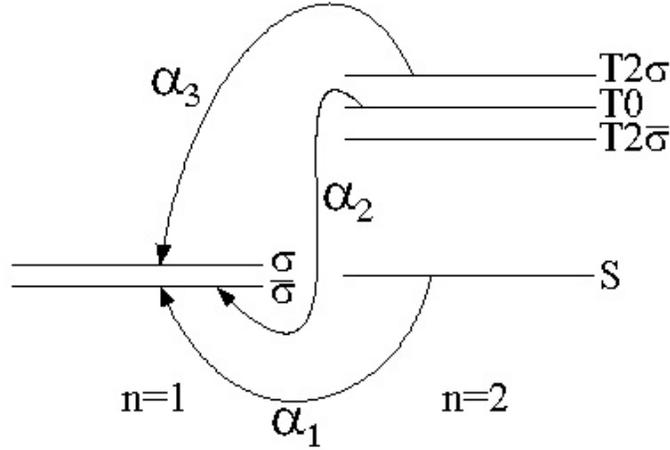}
\end{center}
\caption{\label{fig2}Quasiparticle excitations between state in
effective singlet-triplet model (excitations for only one spin
projections are shown).}
\end{figure}

Processes described by Hamiltonian (\ref{eq5}) are shown in the
figure \ref{fig2}. We have three Fermi-type excitations and hence
three base root vectors {$\alpha_{1}$, $\alpha_{2}$,
$\alpha_{3}$}. Corresponding basis of the Hubbard operators is:
\begin{equation}
\left\{ X_{i}^{\bar{\sigma} S}, X_{i}^{\bar{\sigma} \ T0},
X_{i}^{\sigma \ T2\sigma} \right\} \Rightarrow \left\{ X_{i}^{1},
X_{i}^{2}, X_{i}^{3} \right\}.
\end{equation}
The condition of basis completeness has a form:
\begin{equation}
\sum_{\sigma} X_{f}^{\sigma \sigma} + X_{f}^{S S} + \sum_{M}
X_{f}^{TM \ TM}=1,
\end{equation}
where $M=\left\{ 2\sigma, \ 0, \ 2\bar{\sigma} \right\}$.

In the introduced notations $H_{t}$ has the following form:
\begin{equation}
H_{t} = \sum_{<i,j>, \sigma} \sum_{m,n} \gamma_{i j,
\sigma}(m,n)\,{X_{i \sigma}^{m}}^{+} X_{j\sigma}^{n}, \label{eq8}
\end{equation}
where $m$ and $n$ are root vectors. Also, for further
convenience, direct dependence on spin $\sigma$ is introduced.
Matrix $\gamma_{ij,\,\sigma}(m,n)$ can easily be obtained from
Hamiltonian (\ref{eq5}):
\begin{equation}
\gamma_{ij,\sigma} (m,n)=\left(
\begin{array}{ccc}
t_{ij}^{bb}  & \frac{1}{\sqrt{2} } \gamma_{b}  t_{ij}^{ab}  &
-2\sigma  \gamma_{b}  t_{ij}^{ab}  \\
\frac{1}{\sqrt{2} } \gamma_{b}  t_{ij}^{ab}  & \frac{1}{2}
t_{ij}^{aa}  & -\sigma \sqrt{2}  t_{ij}^{aa}  \\
-2\sigma  \gamma_{b}  t_{ij}^{ab}  & -\sigma \sqrt{2}
 t_{ij}^{aa}  & t_{ij}^{aa}
\end{array}
\right),
\end{equation}

Equations of motion for Hamiltonian (\ref{eq5}) can be expressed
as:
\begin{equation}
\rmi \frac{\rmd}{\rmd t} X_{f\sigma}^{p} =\left[ X_{f\sigma}^{p}
,H_{0} +H_{t} +H_{J} \right] =\Omega_{p} X_{f\sigma}^{p}
+L_{f\sigma}^{p}, \label{eq9}
\end{equation}
where $ L_{f\sigma}^{p} = L_{f\sigma}^{p}(t)+L_{f\sigma}^{p}(J)$,
$L_{f\sigma}^{p}(t) \equiv [X_{f\sigma}^{p},H_{t}]$,
$L_{f\sigma}^{p}(J) \equiv [X_{f\sigma}^{p},H_{J}]$, and
$\Omega_{p} \equiv [X_{f\sigma}^{p},H_{0}]$ is vector of
one-electronic energies:
\begin{equation}
\Omega_{p}=\left(
\begin{array}{l}
\varepsilon_{2S}-\varepsilon_{1}  \\
\varepsilon_{2T}-\varepsilon_{1}  \\
\varepsilon_{2T}-\varepsilon_{1}
\end{array}
\right).
\end{equation}

Taking into account equation (\ref{eq8}) and identity $\left[
A,BC\right] \equiv \left\{ A,B\right\} C-B\left\{ A,C\right\}$ we
can right away get an expression for $L_{f\sigma}^{p}(t)$:
\begin{equation}
L_{f\sigma}^{p}(t) = \sum_{g, \sigma'} \sum_{m,n} \gamma_{fg,
\sigma'}(m,n)\left( E_{f}^{\sigma \sigma'}(p,m) X_{g\sigma'}^{n}
- {X_{g \sigma'}^{m}}^{+} D_{f}^{\sigma \sigma'}(p,n) \right),
\end{equation}
where $E_{f}^{\sigma \sigma'}(p,m) \equiv \left\{
X_{f\sigma}^{p}, {X_{f\sigma'}^{m}}^{+} \right\}$ is the neutral
boson and $D_{f}^{\sigma \sigma'}(p,n) \equiv \left\{
X_{f\sigma}^{p}, X_{f\sigma'}^{n} \right\}$ is $2e$ charged boson:

\begin{eqnarray}
\fl E_{f}^{\sigma \sigma'}(p,m) = \delta_{\sigma \sigma '}
\left(\begin{array}{ccc} X_{f}^{SS} +X_{f}^{\bar{\sigma}
\bar{\sigma} } & X_{f}^{S T0} & 0 \\
X_{f}^{T0 S}  & X_{f}^{T0 T0} + X_{f}^{\bar{\sigma} \bar{\sigma}
} & 0 \\
0 & 0 & X_{f}^{T2\sigma  T2\sigma} + X_{f}^{\sigma \sigma}
\end{array}
\right) \nonumber \\
\lo + \delta_{\bar{\sigma} \sigma '} \left(\begin{array}{ccc}
X_{f}^{\bar{\sigma} \sigma} & 0 & X_{f}^{S T2\sigma}  \\
0 & X_{f}^{\bar{\sigma} \sigma} & X_{f}^{T0 T2\sigma}  \\
X_{f}^{T2\bar{\sigma} S} & X_{f}^{T2\bar{\sigma} T0} & 0
\end{array}
\right), \\
\fl D_{f}^{\sigma \sigma '}(p,n)=0.
\end{eqnarray}

Vanishing of $D_{f}^{\sigma \sigma'}(p,n)$ is due to neglect of
two-hole excitations $| 0 \rangle \rightarrow | S \rangle $ in the
t-t'-J model. Nevertheless such excitations had their effect
resulting in the superexchange interaction J.

It's easy to rewrite expression for $L_{f\sigma}^{p}(t)$ in
reciprocal space (or so-called k-space):
\begin{equation}
L_{k\sigma}^{p}(t) = \sum_{q,\sigma'} \sum_{m,n} \gamma_{-q,
\sigma'}(m,n) E_{k-q}^{\sigma \sigma'}(p,m) X_{q \sigma'}^{n}.
\label{eq10}
\end{equation}

Introducing for convenience operator:
\begin{equation}
Y_{q \sigma} \equiv \frac{1}{2} J_{-q} X_{q}^{\sigma \sigma} +
\delta J_{-q} X_{q}^{\bar{\sigma} \bar{\sigma}}, \label{eq11}
\end{equation}
we can now write down $L_{f\sigma}^{p}(J)$:
\begin{eqnarray}
\fl L_{k\sigma}^{1}(J) = \frac{1}{2} \sum_{q} \left( X_{k-q}^{\bar{\sigma} S} Y_{q\sigma} + Y_{q\sigma} X_{k-q}^{\bar{\sigma} S} - \frac{1}{2} J_{-q} \left( X_{k-q}^{\sigma S} X_{q}^{\bar{\sigma} \sigma} + X_{q}^{\bar{\sigma} \sigma} X_{k-q}^{\sigma S} \right) \right), \\
\fl L_{k\sigma}^{2}(J) = \frac{1}{2} \sum_{q} \left( X_{k-q}^{\bar{\sigma} \ T0} Y_{q\sigma} + Y_{q\sigma} X_{k-q}^{\bar{\sigma} \ T0} - \frac{1}{2} J_{-q} \left( X_{k-q}^{\sigma \ T0} X_{q}^{\bar{\sigma} \sigma} + X_{q}^{\bar{\sigma} \sigma} X_{k-q}^{\sigma \ T0} \right) \right), \\
\fl L_{k\sigma}^{3}(J) = \frac{1}{2} \sum_{q}\left(
X_{k-q}^{\sigma \ T2\sigma} Y_{q\bar{\sigma}} + Y_{q
\bar{\sigma}} X_{k-q}^{\sigma \ T2\sigma} - \frac{1}{2} J_{-q}
\left( X_{k-q}^{\bar{\sigma} \ T2\sigma} X_{q}^{\sigma
\bar{\sigma}} + X_{q}^{\sigma \bar{\sigma}} X_{k-q}^{\bar{\sigma}
\ T2\sigma} \right) \right).
\end{eqnarray}

For decoupling of equations on Green functions we will use the
method of irreducible operators \cite{bib17}. This method based
on linearization of equations of motion within Generalized
Hartree-Fock Approximation (GHFA). Let us first introduce the
irreducible operator:
\begin{equation}
\overline{L_{k\sigma}^{p}} = L_{k\sigma}^{p} - \sum_{h}
C_{k\sigma}(p,h) X_{k\sigma}^{h} - \sum_{h} \Delta_{k\sigma}(p,h)
{X_{-k\bar{\sigma}}^{h}}^{+}, \label{eq12}
\end{equation}
where
\begin{eqnarray}
C_{k\sigma}(p,h) = \frac{\left\langle \left\{ L_{k\sigma}^{p}, {X_{k\sigma}^{h}}^{+} \right\} \right\rangle } {\left\langle \left\{X_{k\sigma}^{h}, {X_{k\sigma}^{h}}^{+} \right\} \right\rangle }, \\
\Delta_{k\sigma}(p,h) = \frac{\left\langle \left\{
L_{k\sigma}^{p}, X_{-k\bar{\sigma}}^{h} \right\} \right\rangle }
{\left\langle \left\{{X_{-k\bar{\sigma}}^{h}}^{+},
X_{-k,\bar{\sigma}}^{h} \right\} \right\rangle }. \label{eq13}
\end{eqnarray}

Here $\Delta_{k\sigma}$ connected to the superconducting order
parameter. Its symmetrical properties will be analyzed in next
chapter.

Now we can write down expressions for $C_{k\sigma}(p,h)$ and
$\Delta_{k\sigma}(p,h)$ in compact form:
\begin{eqnarray}
\fl C_{k\sigma}(p,h) = \frac{1}{\left\langle E_{k=0}^{\sigma \sigma}(h,h) \right\rangle}
\sum_{q} \left\langle E_{q}^{\sigma \sigma}(h,h) \right\rangle \nonumber \\
\lo \times \left[ \gamma_{q-k,\sigma}(p,h)
\left\langle E_{-q}^{\sigma \sigma}(p,p) \right\rangle +
\delta_{p,h} \left\langle Y_{-q\sigma}(p,h) \right\rangle \right], \label{eq14} \\
\fl \Delta_{k\sigma}(p,h) = \frac{1}{\left\langle
E_{k=0}^{\bar{\sigma} \bar{\sigma}}(h,h) \right\rangle} \sum_{q}
\left[ - A_{q-k}^{(1)}(p,h)
B_{q\sigma}(h,p) - A_{q+k}^{(2)}(p,h) B_{q\sigma}(p,h) \right. \nonumber \\
\lo + \left. \sum_{m,n} \left( \gamma_{-q,\sigma}(m,n)
R_{q\sigma}^{(1)}(p,m;h,n) - \gamma_{-q,\bar{\sigma}}(m,n)
R_{q\sigma}^{+(2)}(p,m;h,n) \right) \right]. \label{eq15}
\end{eqnarray}
Tensors $R_{q\sigma}^{(1)} (p,m;h,n)$ and $R_{q\sigma}^{+\,(2)}
(p,m;h,n)$ are presented in \ref{appendixA}.

Averaging-out in $C_{k\sigma}(p,h)$ were made in Hubbard-I
approximation and all averages in $C_{k\sigma}(p,h)$ are as
follows:
\begin{eqnarray}
\fl \left\langle E_{k}^{\sigma \sigma'}(p,h) \right\rangle =
\delta_{\sigma \sigma'} \left(
\begin{array}{ccc}
\left\langle X_{k}^{SS} +X_{k}^{\bar{\sigma} \bar{\sigma} }
\right\rangle  & 0 & 0 \\
0 & \left\langle X_{k}^{T0 T0} +X_{k}^{\bar{\sigma} \bar{\sigma
} } \right\rangle  & 0 \\
0 & 0 & \left\langle X_{k}^{T2\sigma  T2\sigma} +X_{k}^{\sigma
\sigma} \right\rangle
\end{array}
\right), \label{eq16} \\
\fl \left\langle Y_{q\sigma}(p,h) \right\rangle = \left(
\begin{array}{ccc}
\left\langle Y_{q\sigma} \right\rangle  & 0 & 0 \\
0 & \left\langle Y_{q\sigma} \right\rangle  & 0 \\
0 & 0 & \left\langle Y_{q\bar{\sigma} } \right\rangle
\end{array}
\right), \label{eq17}
\end{eqnarray}
where $Y_{q\sigma}$ is determined by (\ref{eq11}). Also the
anomalous averages were introduced:
\begin{equation}
B_{k\sigma}(p,h) \equiv \left\langle X_{-k\bar{\sigma}}^{p}
X_{k\sigma}^{h} \right\rangle. \label{eq18}
\end{equation}
Direct calculations can reveal one useful property of this
averages:
\begin{equation}
B_{-k\bar{\sigma}}(h,p) = -B_{k\sigma}(p,h). \label{eq19}
\end{equation}

At calculation of $ \left\langle \left\{ L_{k\sigma}^{3}(J),
X_{-k\bar{\sigma}}^{h} \right\} \right\rangle $ the terms of type
$ \left\langle X_{k-q}^{\bar{\sigma} \ T2\sigma}
P_{k,q}^{\sigma}(h) \right\rangle$ appears where $ P_{k,
q}^{\sigma}(h) = \sqrt{N} [X_{q}^{\bar{\sigma} \sigma},
X_{-k\bar{\sigma}}^{h}]$ is a Fermi-like operator. These terms are
responsible for 3-fermion excitations and in the Hubbard-I
approximation they are equal to zero.

Tensors $R_{q\sigma}^{(1)}(p,m;h,n)$ and
$R_{q\sigma}^{+\,(2)}(p,m;h,n)$ are presented in Appendix I.
Shortly, each of their elements is equal to anomalous average
$B_{q\sigma}$ or zero. Matrixes $A_{q-k}^{(1)}(p,h)$ and
$A_{q+k}^{(2)}(p,h)$ consist of exchange integrals and has the
following explicit form:
\begin{eqnarray}
A_{q-k}^{(1)}(p,h) = \frac{1}{2} \left(
\begin{array}{ccc}
J_{q-k}  & J_{q-k}  & 2\delta J_{q-k}  \\
J_{q-k}  & J_{q-k}  & 2\delta J_{q-k}  \\
2\delta J_{q-k}  & 2\delta J_{q-k}  & 2\delta J_{q-k}
\end{array}
\right), \\
A_{q+k}^{(2)}(p,h) = \frac{1}{2} \left(
\begin{array}{ccc}
J_{q+k}  & J_{q+k}  & 0 \\
J_{q+k}  & J_{q+k}  & 0 \\
0 & 0 & 0
\end{array}
\right).
\end{eqnarray}

We can now perform elementary check of the obtained
$\Delta_{k\sigma}(p,h)$ and $C_{k\sigma}(p,h)$ by using fact that
when moving energy of the triplet $TM$ to the infinity we have to
obtain the usual t-J model \cite{bib18,bib19}. Because in paper
\cite{bib18} Hubbard operators were also used we will compare our
gap and renormalization of spectrum with its results:
\begin{eqnarray}
C_{k\sigma}^{t-J} = t_{k} (1-n_{\bar{\sigma}}) - \frac{J}{2} n_{\bar{\sigma}}, \\
\Delta_{k\sigma}^{t-J} = \frac{1}{1-n_{\sigma}} \sum_{q} \left(
2t_{q} - \frac{1}{2} \left( J_{k+q} + J_{k-q} \right) \right)
B_{q\sigma}^{t-J}.
\end{eqnarray}
Neglecting all excitations to the triplet zone ($\alpha_{2}
\rightarrow 0$, $\alpha_{3} \rightarrow 0$, $\delta J_{q}
\rightarrow 0$, and $t_{fg}^{nm} \rightarrow 0$ if n or m is
equal to a) coefficients and matrixes in singlet-triplet model
becomes:
\begin{eqnarray}
\gamma_{fg,\sigma}(m,n) = \delta_{m 1} \delta_{n 1} t_{fg}^{bb}, \\
\left\langle E_{f}^{\sigma \sigma'} \right\rangle = \delta_{p 1} \delta_{m 1} \delta_{\sigma, \sigma'} \left\langle X_{f}^{S S} + X_{f}^{\bar{\sigma} \bar{\sigma}} \right\rangle, \\
B_{q\sigma}(p,h) = \delta_{p 1} \delta_{h 1} \left\langle X_{-q\bar{\sigma}}^{1} X_{q\sigma}^{1} \right\rangle \equiv B_{q\sigma}, \\
\left\langle Y_{q\sigma}(p,h) \right\rangle = \delta_{p 1}
\delta_{h 1} \left\langle \frac{1}{2} J_{-q} X_{q}^{\sigma \sigma}
\right\rangle.
\end{eqnarray}
Having made all these limiting transitions in (\ref{eq14}) and
(\ref{eq15}) we derive:
\begin{eqnarray}
\fl C_{k\sigma} \rightarrow \frac{1}{1-n_{\bar{\sigma}}} \sum_{q}(1-n_{\bar{\sigma}})
\left( t_{q-k}^{bb} (1-n_{\bar{\sigma}}) +
\frac{1}{2} J_{-q} (-n_{\bar{\sigma}})\right) \delta_{q,0} \nonumber \\
\lo = t_{k}^{bb} (1-n_{\bar{\sigma}}) - \frac{1}{2} J_{0} n_{\bar{\sigma}}, \\
\fl \Delta_{k\sigma} \rightarrow \frac{1} {1-n_{\sigma}} \sum_{q}
\left( 2t_{-q}^{bb} - \frac{1}{2} \left( J_{q-k} + J_{q+k}
\right) \right) B_{q\sigma}.
\end{eqnarray}
This is exactly what we have in the t-J model. So the effective
singlet-triplet model in the low energy limit (triplet energy
tend to infinity) reduces to the t-J model.

Let's return to decoupling of Green functions. In GHFA we have to
put irreducible operator $\overline{L_{k\sigma}^{p}}$ to zero. In
this case the equations of motion take the following form:
\begin{equation}
\fl \cases{ \rmi \frac{\rmd}{\rmd t} X_{k\sigma}^{p} = \sum_{h}
\left[ \Omega_{p} \ \delta_{p,h} + C_{k\sigma}(p,h) \right]
X_{k\sigma}^{h} +
\sum_{h} \Delta_{k\sigma}(p,h) {X_{-k\bar{\sigma}}^{h}}^{+} \\
\rmi \frac{\rmd}{\rmd t} {X_{-k\bar{\sigma}}^{p}}^{+} = -
\sum_{h} \left[ \Omega_{p} \ \delta _{p,h} +
C_{-k\bar{\sigma}}^{+}(p,h) \right] {X_{-k\bar{\sigma}}^{h}}^{+} -
\sum_{h} \Delta_{-k\bar{\sigma}}^{+}(p,h) X_{k\sigma}^{h} }
\label{eq20}
\end{equation}
Now we can easily write down a system of Gorkov type equations
for normal and abnormal, and this system is closed:
\begin{equation}
\fl \cases{ \sum_{h} \left[ \left( E-\Omega_{p} \right)
\delta_{p,h} - C_{k\sigma}(p,h) \right] \hat{G}_{k\sigma}(h,l) -
\sum_{h} \Delta_{k\sigma} (p,h) \hat{F}_{k\sigma}^{+}(h,l) =
\left\langle \left\{ X_{k\sigma}^{p}, {X_{k\sigma}^{l}}^{+} \right\} \right\rangle \\
\sum_{h} \left[ \left( E+\Omega_{p} \right) \delta_{p,h} +
C_{-k\bar{\sigma}}^{+}(p,h) \right] \hat{F}_{k\sigma}^{+}(h,l) +
\sum_{h} \Delta_{-k\bar{\sigma}}^{+}(p,h) \hat{G}_{k\sigma}(h,l)
= 0 }
\end{equation}
Actually this is an equation for a matrix normal Green function
$\hat{G}_{k\sigma}=\left\langle \left\langle X_{k\sigma}^{h}
\right.\left| {X_{k\sigma}^{l}}^{+} \right\rangle \right\rangle$
and matrix abnormal Green function
$\hat{F}_{k\sigma}^{+}=\left\langle \left\langle
{X_{-k\bar{\sigma}}^{h}}^{+} \right.\left| {X_{k\sigma}^{l}}^{+}
\right\rangle \right\rangle$.

Introducing matrixes:
\begin{eqnarray}
\hat{\Re}_{k\sigma} = \left( \Omega_{p} \delta_{p,h} + C_{k\sigma}(p,h)\right), \\
\hat{\Delta}_{k\sigma} = \Delta_{k\sigma}(p,h), \\
\hat{E}_{k}^{\sigma \sigma} = E_{k}^{\sigma \sigma}(p,h) \equiv
\sqrt{N} \left\{ X_{k\sigma}^{p}, {X_{k\sigma}^{l}}^{+} \right\},
\end{eqnarray}
and solving equations we obtain expressions for
$\hat{G}_{k\sigma}$ and $\hat{F}_{k\sigma}^{+}$:
\begin{equation}
\cases{ \hat{F}_{k\sigma}^{+} = - \left[ E\hat{I} +
\hat{\Re}_{-k\bar{\sigma}}^{+}
\right]^{-1} \hat{\Delta}_{-k\bar{\sigma}}^{+} \hat{G}_{k\sigma}  \\
\hat{G}_{k\sigma} = \left[ E\hat{I} -\hat{\Re }_{k\sigma}
+\hat{\Delta }_{k\sigma} \left( E\hat{I} +
\hat{\Re}_{-k\bar{\sigma}}^{+} \right) ^{-1}
\hat{\Delta}_{-k\bar{\sigma}}^{+} \right]^{-1} \left\langle
\hat{E}_{k}^{\sigma \sigma} \right\rangle / \sqrt{N} }
\label{eq21}
\end{equation}
where $\hat{I}$ is identity matrix and N is number of vectors in
k-space.

It can be seen now by analogy with the BCS theory of
low-temperature superconductivity that $\Delta_{k\sigma}$ is the
superconducting order parameter.

The system (\ref{eq21}) is the set of matrix Green function and
can be used to obtain energy spectrum and averages for any
problem with defined basis of root vectors. In following chapters
we will use system (\ref{eq21}) to perform symmetry
classification of order parameter and to investigate normal
paramagnetic phase.

\section{Symmetry Classification of Superconducting Order Parameter}

An early suggestion that AF spin fluctuation could give rise to
singlet $d_{x^{2}-y^{2}}$- wave pairing in p-type cuprate
superconductors was made by Bickers, Scalapino and Scalettar
\cite{bib20}. This suggestion has been supported by the FLEX
approximation to the Hubbard model \cite{bib21} that is, however
not valid in the case of $U \gg t$. In this limit of SCES the
proper model is the t-J model. Exact diagonalization and quantum
Monte-Carlo method results for small clusters have been discussed
by Dagotto \cite{bib22}. For the infinite lattice the most
adequate perturbation approach to the t-J model has been
formulated in the X-operator representation because of the exact
treatment of local constraint due to X-operators algebra. The
mean-field solution \cite{bib23,bib24,bib25} of the t-J model and
analysis of the self-energy correlations beyond the mean-field
approximation by diagram technique \cite{bib26} and by high-order
decoupling scheme \cite{bib19} has confirmed the
$d_{x^{2}-y^{2}}$- pairing in the hole-doped system with typical
$T_{c}(x)$ dependence. Latest experimental results appear to
prove $d_{x^{2}-y^{2}}$- pairing not only in p-type systems but
also in n-type systems (see review \cite{bib27}).

\begin{figure}[h]
\begin{center}
\includegraphics[width=0.7\textwidth]{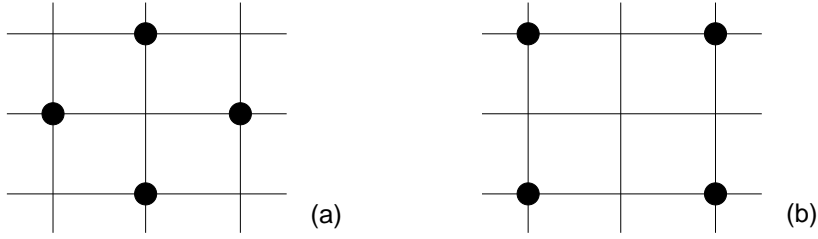}
\end{center}
\caption{\label{fig3}Location of nearest neighbors (a) and
next-nearest neighbors (b) on square lattice.}
\end{figure}

Let's proceed to a symmetry classification of the order parameter
$\Delta_{k\sigma}(p,h)$ in the effective singlet-triplet model
considering case of square lattice. First, we have to break
hopping and exchange integrals in two terms:
\begin{eqnarray}
& & \gamma_{k}(p,h) = t(p,h) \omega_{k} + t'(p,h) \tilde{\omega}_{k}, \\
& & J_{k} = J \omega_{k} + J' \tilde{\omega}_{k}, \\
& & \delta J_{k} = \delta J \omega_{k} + \delta J'
\tilde{\omega}_{k},
\end{eqnarray}
where
\begin{eqnarray}
\omega_{k} & = & \sum_{\delta} \exp{\left(\rmi k\delta \right)} = 2\left( \cos k_{x} + \cos k_{y} \right), \\
\tilde{\omega }_{k} & = & \sum_{\delta'} \exp{\left( \rmi k
\delta' \right)} = 4 \cos k_{x} \cos k_{y},
\end{eqnarray}
and non-primed values are concerned with nearest neighbor (figure
\ref{fig3}(a), first coordination sphere) and primed values are
concerned with next-nearest neighbor (figure \ref{fig3}(b), second
coordination sphere).

Second, to make a classification we will distinguish following
symmetry types:

\subsection{s-type}

In superconducting phase we have a constraint condition:
\begin{equation}
\frac{1}{N} \sum_{k} \left\langle a_{k\sigma} a_{-k\bar{\sigma}}
\right \rangle = 0. \label{eq22}
\end{equation}
Right side of this identity in the effective singlet-triplet
model is equal to zero due to choose of  basis of root vectors
$\alpha_{i}$, i.e. because of absence of the transitions from
lower to higher Hubbard bands. Moreover, condition (\ref{eq22})
is satisfied only in case of p- and d-pairing but not for
symmetric s-pairing. Hence in the singlet-triplet model the
symmetric s-type singlet pairing is absent.

\subsection{p-type}

Triplet pairing of p-type is impossible because for realization
of this symmetry there must be a ferromagnetic interaction (see
e.g. \cite{bib18}) but in case of singlet-triplet model we have
only antiferromagnetic exchange.

\subsection{d-type}

Singlet d-type pairing is forbidden neither by the constraint
condition (\ref{eq22}) nor by the type of interaction.

First consider $d_{x^{2}-y^{2}}$- pairing type. In this case
there is a restriction:
\begin{equation}
\sum_{q} \cos q_{x} B_{q\sigma}(p,h) = -\sum_{q} \cos q_{y}
B_{q\sigma}(p,h), \label{eq23}
\end{equation}
and, as a consequence,
\begin{equation}
\sum_{q} \sin q_{x} B_{q} = 0,
\end{equation}
and
\begin{equation}
\sum_{q} \sin q_{y} B_{q} = 0.
\end{equation}

This immediately leads to the significant simplification of
superconducting gap $\Delta_{k\sigma}(p,h)\equiv
\Delta_{k\sigma}^{(d_{x^{2} -y^{2}})}(p,h)$:
\begin{equation}
\Delta_{k\sigma}^{(d_{x^{2}-y^{2}})}(p,h) = -\frac{2
\Delta_{\sigma}^{(d_{x^{2}-y^{2}})}(p,h)} {\sqrt{N} \left\langle
E_{k=0}^{\bar{\sigma} \bar{\sigma}}(h,h) \right\rangle} (\cos
k_{x} - \cos k_{y}). \label{eq24}
\end{equation}
where $\Delta_{\sigma}^{(d_{x^{2}-y^{2}})}(p,h)$ is the
impulse-independent part of the gap:
\begin{equation}
\Delta_{\sigma}^{(d_{x^{2}-y^{2}})} \equiv \left(
\begin{array}{ccc}
J \Psi_{ \sigma}^{1,1}  & J\left( \Psi_{ \sigma}^{1,2} +\Psi_{
\sigma}^{2,1} \right) /2 & \delta
J \Psi_{ \sigma}^{3,1}  \\
J\left( \Psi_{ \sigma}^{1,2} +\Psi_{ \sigma}^{2,1} \right) /2 & J
\Psi_{ \sigma}^{2,2}  & \delta
J \Psi_{ \sigma}^{3,2}  \\
\delta J \Psi_{ \sigma}^{1,3}  & \delta J \Psi_{ \sigma}^{2,3}  &
\delta J \Psi_{ \sigma}^{3,3}
\end{array}
\right), \label{eq25}
\end{equation}

\begin{equation}
\Psi_{\sigma}^{p,h} \equiv \sum_{q} (\cos q_{x} - \cos q_{y})
B_{q\sigma}(p,h). \label{eq26}
\end{equation}
Impulse-independent $\Delta_{\sigma}^{(d_{x^{2}-y^{2}})}(p,h)$
includes exchange integrals only for nearest neighbors and
therefore can exist only in boundaries of first coordination
sphere.

Now lets consider $d_{xy}$- pairing type. The restriction made by
this symmetry is as follows:
\begin{equation}
\sum_{q} \cos q_{x} \cos q_{y} B_{q\sigma}(p,h) = 0. \label{eq27}
\end{equation}
Also
\begin{equation}
\sum_{q} \sin q_{x} \cos q_{y} B_{q} = 0,
\end{equation}
and
\begin{equation}
\sum_{q} \cos q_{x} \sin q_{y} B_{q} = 0.
\end{equation}

The superconducting gap $\Delta_{k\sigma}(p,h) \equiv
\Delta_{k\sigma}^{(d_{xy})}(p,h)$ takes the form:
\begin{equation}
\Delta_{k\sigma}^{(d_{xy})}(p,h) = -\frac{4
\Delta_{\sigma}^{(d_{xy})}(p,h)} {\sqrt{N} \left\langle
E_{k=0}^{\bar{\sigma} \bar{\sigma}}(h,h) \right\rangle} \sin
k_{x} \sin k_{y}. \label{eq28}
\end{equation}
where impulse-independent part of the gap
$\Delta_{\sigma}^{(d_{xy})}(p,h)$:
\begin{equation}
\Delta_{\sigma}^{(d_{xy})} \equiv \left(
\begin{array}{ccc}
J' \Phi_{ \sigma}^{1,1}  & J'\left( \Phi _{ \sigma}^{1,2} +\Phi_{
\sigma}^{2,1} \right) /2 &
\delta J' \Phi_{ \sigma}^{3,1}  \\
J'\left( \Phi_{ \sigma}^{1,2} +\Phi_{ \sigma }^{2,1} \right) /2 &
J' \Phi_{ \sigma}^{2,2}  &
\delta J' \Phi_{ \sigma}^{3,2}  \\
\delta J' \Phi_{ \sigma}^{1,3}  & \delta J' \Phi_{ \sigma}^{2,3}
& \delta J' \Phi _{ \sigma}^{3,3}
\end{array}
\right), \label{eq29}
\end{equation}

\begin{equation}
\Phi_{\sigma}^{p,h} \equiv \sum_{q} \sin q_{x} \sin q_{y}
B_{q\sigma}(p,h). \label{eq30}
\end{equation}

Coexistence of $d_{x^{2}-y^{2}}$- and $d_{xy}$- pairing types is
forbidden by the group theory: in the considered case of
tetragonal lattice symmetry $d_{x^{2}-y^{2}}$- and $d_{xy}$-
types belongs to different irreducible representation (see review
\cite{bib27}) so there must be concurrence between these types of
pairing.

Note that triplet channel results in $\delta J$ in matrix
(\ref{eq25}) and $\delta J'$ in matrix (\ref{eq29}). Obviously
triplet channel gives the additional pairing.

It is very interesting that $d_{xy}$-type includes exchange
integrals only for next-nearest neighbors (see equation
(\ref{eq29})) and therefore can exist only in second coordination
sphere. Hence in nearest neighbor approximation we can have only
order parameter of $d_{x^{2}-y^{2}}$-symmetry.

Actually equations (\ref{eq24}) and (\ref{eq28}) are equations
for order parameters $\Psi_{\sigma}^{p,h}$ and
$\Phi_{\sigma}^{p,h}$ respectively and should be solved
self-consistently via Green function equations (\ref{eq21}). When
this is done we can obtain energy spectrum and phase diagram
$T_{c}$ versus concentration $x$.

\subsection{Other symmetry types}

The other symmetry types are not realized due to absence of
corresponding combinations of trigonometrical functions in
expression for superconducting gap (\ref{eq15}).

\bigskip

Combination of previously analyzed types such as s+d is
impossible because in the considered case of tetragonal lattice
symmetry s- and d- types belongs to different irreducible
representation (see review \cite{bib27} and references from
there). So the order parameter symmetry must be of d-type or,
more precisely, to be a concurrence of $d_{x^{2}-y^{2}}$ and
$d_{xy}$ singlet pairing types. This conclusion is true not only
for hole doped systems but also for electron doped systems
because in the limit of infinite energy of singlet-triplet
excitation (i.e. absence of excitations to the triplet states -
this case corresponds to n-type cuprates) our equations take the
form of t-J model ones and the gap symmetry remains.

\section{Normal Paramagnetic Phase}

In normal paramagnetic phase $\Delta_{k\sigma}=0$ and Green
function equations (\ref{eq21}) become essentially simpler:
\begin{equation}
\cases{ \hat{G}_{k\sigma} = \left( E\hat{I} -\hat{\Re }_{k\sigma}
\right)^{-1} \left\langle \hat{E}_{k}^{\sigma
\sigma} \right\rangle / \sqrt{N} \\
\hat{F}_{k\sigma}^{+} = 0 }
\end{equation}

By solving these equations one can obtain energy spectrum and
self-consistently find Fermi level. Also using following
definition for the spectral density in terms of one-electron
annihilation operators $c_{k}^{+}$:
\begin{equation}
A(k,\sigma,E) = -\frac{1}{\pi} \mathrm{Im} \left\langle
\left\langle c_{k \sigma}^{+} \right. \left| c_{k \sigma}
\right\rangle \right\rangle,
\end{equation}
we can calculate density of states (DOS):
\begin{equation}
N(E) = \frac{1}{N} \sum_{k \sigma} A(k,\sigma,E) = \frac{1}{N}
\sum_{k, \sigma, \lambda} \left( -\frac{1}{\pi} \mathrm{Im}
\left\langle \left\langle c_{k \sigma \lambda}^{+} \right.\left|
c_{k \sigma \lambda} \right\rangle \right\rangle \right),
\end{equation}
where $\sigma$ is spin and $\lambda$ is the orbital index. In
terms of Hubbard operators the expression for DOS has the
following explicit form:
\begin{eqnarray}
\fl N(E) = - \frac{1}{N} \sum_{k \sigma} \left[ \left(
\gamma_{x}^{2} + \gamma_{b}^{2} \right) \mathrm{Im} \left\langle
\left\langle {X_{k\sigma}^{1}}^{+} \right.\left| X_{k\sigma}^{1}
\right\rangle \right\rangle
+ \right. \nonumber \\
\lo + \left. \left( \gamma_{z}^{2} + \gamma_{a}^{2} +
\gamma_{p}^{2} \right) \left( \frac{1}{2} \mathrm{Im} \left\langle
\left\langle {X_{k\sigma}^{2}}^{+} \right. \left| X_{k\sigma}^{2}
\right\rangle \right\rangle + \mathrm{Im} \left\langle
\left\langle {X_{k\sigma}^{3}}^{+} \right.\left| X_{k\sigma}^{3}
\right\rangle \right\rangle \right) \right] \nonumber \\
\lo = \frac{1}{N} \sum_{k \sigma} \left[ \left( \gamma_{x}^{2} +
\gamma_{b}^{2} \right) \mathrm{Im}
\hat{G}_{k\sigma}^{\mathrm{a}}(1,1)
+ \right. \nonumber \\
\lo + \left. \left( \gamma_{z}^{2} + \gamma_{a}^{2} +
\gamma_{p}^{2} \right) \left( \frac{1}{2} \mathrm{Im}
\hat{G}_{k\sigma}^{\mathrm{a}}(2,2) + \mathrm{Im}
\hat{G}_{k\sigma}^{\mathrm{a}}(3,3) \right) \right].
\end{eqnarray}
Here $\epsilon$ and $N_{0}(\epsilon)$ is the dispersion and DOS
for non-interacting case, $\gamma_{m}$ is the coefficient of
transformation $c_{k\sigma} = \sum\limits_{m} \gamma_{m}
X_{k\sigma}^{m}$, index "$\mathrm{a}$" of Green function denotes
that $\hat{G}_{k\sigma}^{\mathrm{a}}$ is advance Green function
(while all over the paper we have used retarded Green functions).
There is simple relation between this two types:
$\hat{G}_{k\sigma}(p,h) = - \hat{G}_{k\sigma}^{\mathrm{a}}(h,p)$.

The calculations described above where performed for High-$T_{c}$
superconductor $La_{2-x}Sr_{x}CuO_{4}$ with the following set of
the model parameters (in eV):

\bigskip

\begin{tabular}{lllll}
$t_{pd}=1$, & $\varepsilon(d_{x^{2}-y^{2}})=0$, & $\varepsilon(d_{z^{2}})=2$, & $\varepsilon(p_{x})=1.5$, & $\varepsilon(p_{z})=0.45$, \\
$t_{pp}=0.46$, & $t'_{pp}=0.42$, & $U_{d}=9$, $U_{p}=4$, &
$V_{pd}=1.5$, & $J_{d}=1$.
\end{tabular}

\bigskip

The same parameters where previously used in \cite{bib9.2} for
undoped $La_{2-x}Sr_{x}CuO_{4}$. The energy dispersion, obtained
there, proved to be in good agreement with experimental ARPES
data.

\begin{figure}[h]
\begin{center}
\includegraphics[width=1.0\textwidth]{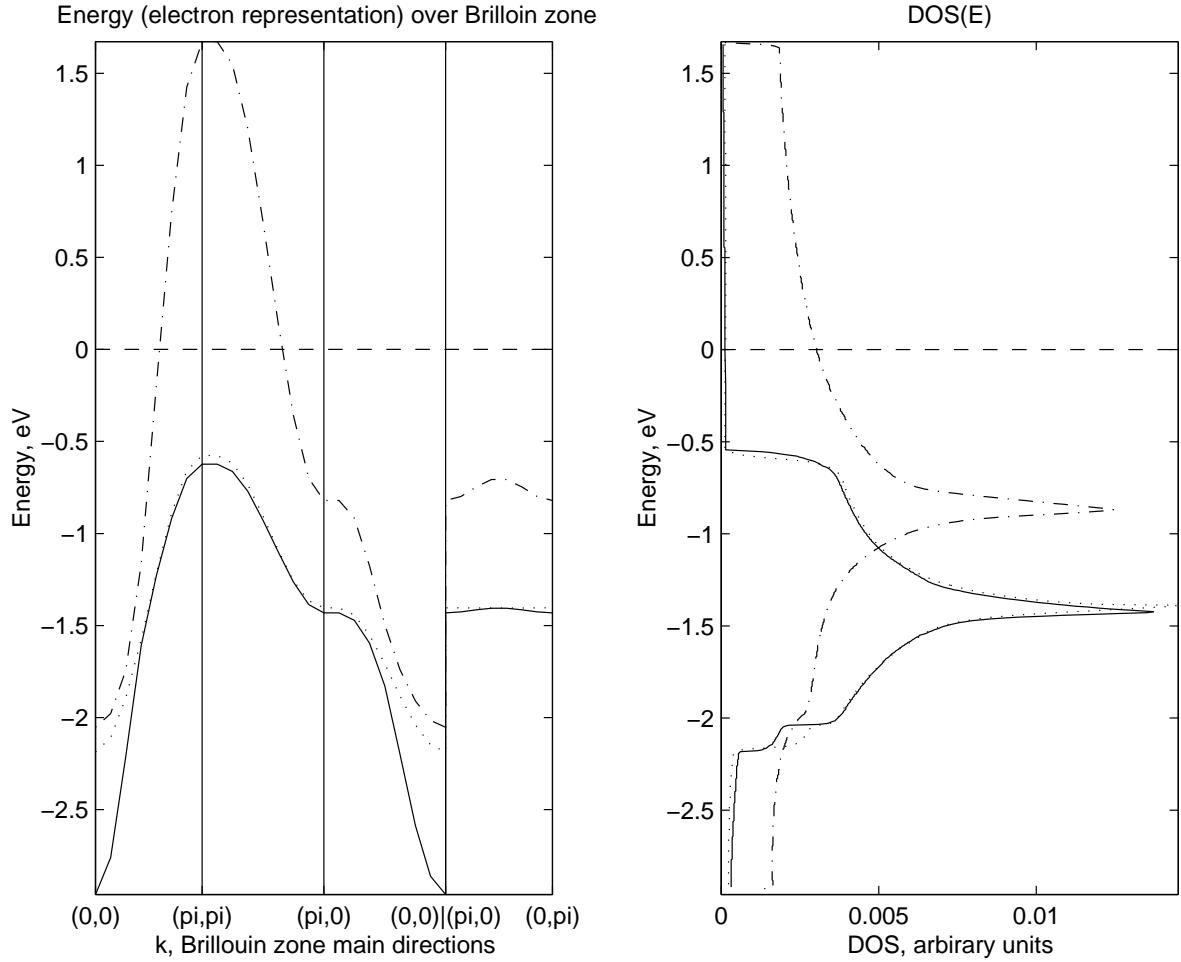}
\end{center}
\caption{\label{fig4}Energy Dispersion (on the left) and Density
of States (on the right) in normal paramagnetic phase of the
effective singlet-triplet model for concentration of dopant
$x=0.2$. Energy of the singlet sub-band and its DOS are marked by
dash-dotted line. Energy of the triplet sub-bands and
corresponding DOS are marked by straight and dotted lines. Dashed
line indicates location of the Fermi level.}
\end{figure}

Our results for slightly overdoped copper oxide (concentration of
dopant $x=0.2$) are presented in figure \ref{fig4}.

\begin{figure}[h]
\begin{center}
\includegraphics[width=1.0\textwidth]{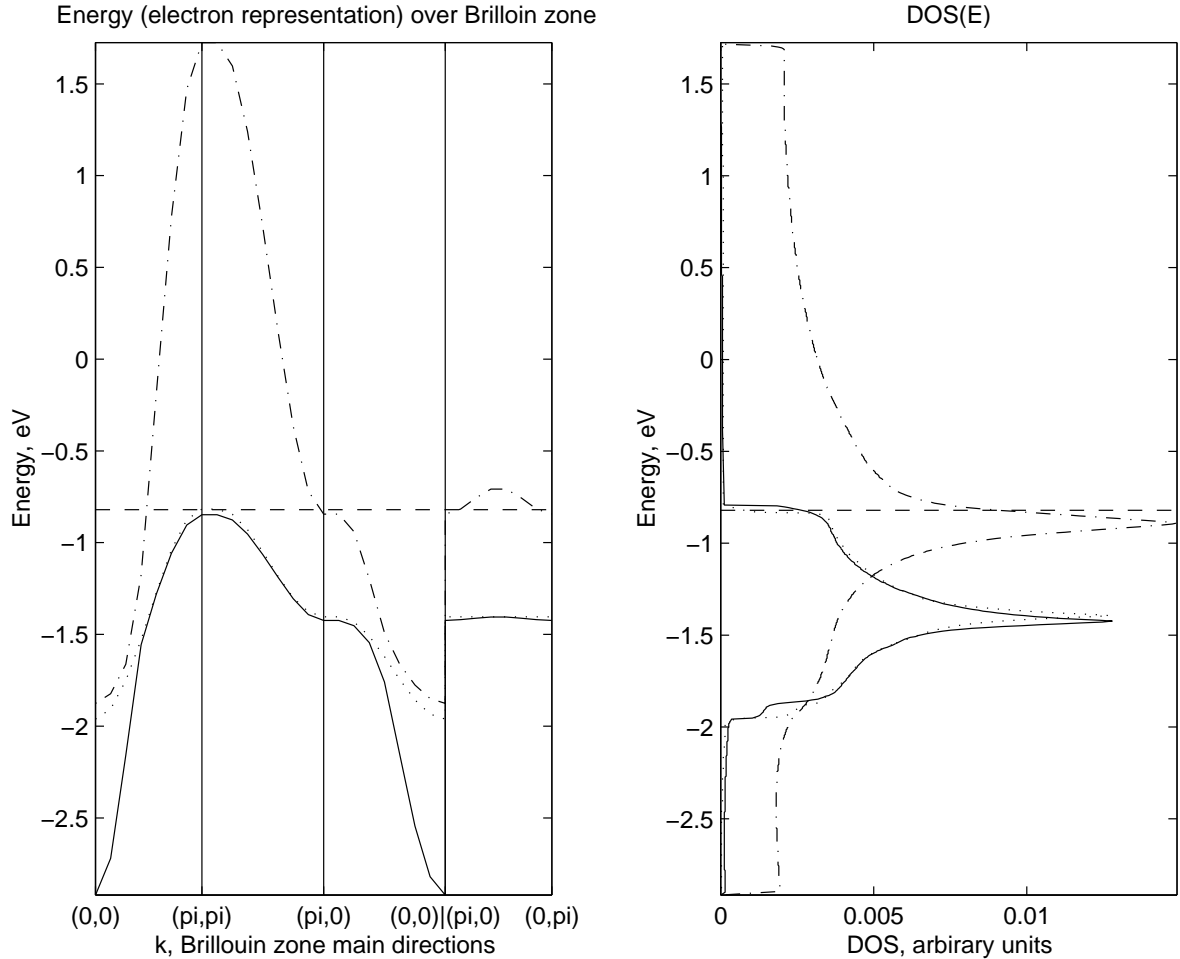}
\end{center}
\caption{\label{fig5}Energy Dispersion (on the left) and Density
of States (on the right) in normal paramagnetic phase of the
effective singlet-triplet model for concentration of dopant
$x=0.43$. All definitions are the same as in figure \ref{fig4}.
Distance between Van-Hove singularity and Fermi level is 0.074 eV}
\end{figure}

Interesting case of coincidence of Fermi level and Van-Hove
singularity is shown in figure \ref{fig5}.

It can be seen that singlet sub-band is very wide - almost 3 eV.
It is a consequence of Hubbard I approximation in spectrum
renormalization term $C_{k\sigma}(p,h)$ (equation (\ref{eq14})).
In more rigorous approximations with inclusion of spin
correlators the energy bands should be more narrow (see
\cite{bib28} and \cite{bib29}).

At the concentration $x=0.43$ Fermi level coincides with Van-Hove
singularity in singlet sub-band. But if we will neglect hopping
on second coordination sphere (nearest neighbor approximation)
then we will get the flattening of dispersion in
$(\pi,0)-(0,\pi)$ direction and shift of Van-Hove singularities
to higher energies. This will result in coincidence of Fermi
level and singularity in singlet sub-band at concentration
$x=0.33$. It is typical value for the t-J model with nearest
neighbor hopping t. It is also known that in t-t'-J model with
the next-nearest neighbor hopping t' the same shift in energies
takes place so this phenomena is common both to t-t'-J model and
the effective singlet-triplet model. Accepting the Van-Hove
scenario of superconductivity, where the optimal doping
correspond to coincident of Fermi level with Van-Hove
singularity, we can clearly see that this should happened at
$x=0.43$. Meanwhile the experimentally obtained optimal doping
value for $La_{2-x}Sr_{x}CuO_{4}$ is $0.18$. So, either the
Van-Hove scenario is not applicable or the model should be
refined. This question could be answered only after the complete
theory of superconductivity in the frame of the effective
singlet-triplet model will be constructed.

\section{Conclusion}

In present work we obtained the effective Hamiltonian of the
singlet-triplet model for copper oxides. Being the generalization
of the t-J model to account of two-particle triplet state
resulting Hamiltonian has several important features. At first,
the $X_{i}^{\sigma \sigma} X_{j}^{\sigma \sigma}$ term appear
which can have a non-trivial contribution to superconducting
pairing. Second, the account of a triplet leads to renormalization
of exchange integral. And third, the singlet-triplet model is
asymmetric for n- and p-type systems. For n-type systems the
usual t-J model takes place while for p-type superconductors with
complicated structure on the top of the valence band the
singlet-triplet transitions plays an important role. The asymmetry
of p- and n-type systems is known experimentally.

The analysis of the possible processes in the effective
singlet-triplet model shows that besides spin-fluctuation
superconducting pairing mechanism typical for the t-J model we
have pairing due to singlet-triplet transitions. These transitions
induce spin-exciton which can play a role of intermediate boson
in superconducting pairing.

We have also performed a symmetry classification of
superconducting order parameter. It was shown that in case of
tetragonal lattice s-type singlet pairing doesn't take place
while the $d_{x^{2}-y^{2}}$- and $d_{xy}$-types of singlet
pairing can exist. Moreover, there must be a concurrence between
$d_{x^{2}-y^{2}}$- and $d_{xy}$-types. At the same time,
$d_{xy}$-type can exist only within the second coordination
sphere and $d_{x^{2}-y^{2}}$-type can exist only within the
first. This fact lets us to take into account only
$d_{x^{2}-y^{2}}$-type symmetry of order parameter in
nearest-neighbors approximation.

As concerns n-type cuprates, the gap symmetry was unclear for a
long time. Recently, phase sensitive tunnel experiments by Tsuei
and Kirtley \cite{bib27} find an evidence for dominant
$d_{x^{2}-y^{2}}$ symmetry in electron doped cuprates. This fact
coincides with our results because in the limit of infinite
energy of singlet-triplet excitation (i.e. absence of excitations
to the triplet states - this case corresponds to n-type cuprates)
our equations take the form of t-J model ones and the
$d_{x^{2}-y^{2}}$-gap symmetry remain.

For normal paramagnetic phase we have obtained the energy
dispersion over Brillouin zone and calculated density of states.
Evolution of Fermi level with doping is also found. Both singlet
and triplet excitations contributes to the density of states which
leads to appearing of two Van-Hove singularities. At holes
concentration $x=0.43$ the Fermi level crosses the first Van-Hove
singularity corresponding to singlet sub-band. And at holes
concentration $x=0.64$ the Fermi level crosses the second
singularity corresponding to triplet sub-band.

\section{Acknowledgements}

The authors would like to thank N.M. Plakida and A.V. Sherman for
discussions and useful comments.

This work has been supported by the RFBR grant 00-02-16110, by
Krasnoyarsk Regional Science Foundation, grant 10F003C, by
RFBR-"Enisey" grant 02-02-97705. INTAS support of research
programme "Electronic and magnetic properties in novel
superconductors: spin fluctuations vs electron-phonon coupling"
at number 654 is also acknowledged.

\clearpage

\appendix
\section{Tensors $R_{q\sigma}^{(1)} (p,m;h,n)$ and $R_{q\sigma}^{+\,(2)}
(p,m;h,n)$} \label{appendixA}

Tensors $R_{q\sigma}^{(1)} (p,m;h,n)$ and $R_{q\sigma}^{+\,(2)}
(p,m;h,n)$ where introduced as follows:
\begin{equation}
\sqrt{N} \left\langle \lbrack X_{-k\bar{\sigma} }^{h},
E_{k-q}^{\sigma \sigma'}(p,m)] X_{q\sigma'}^{n} \right\rangle
\equiv \delta_{\sigma \sigma'} R_{q\sigma}^{(1)}(p,m;h,n) +
\delta_{\bar{\sigma} \sigma'} R_{q\bar{\sigma}}^{(2)}(p,m;h,n).
\end{equation}
The "+" symbol in $R_{q\sigma}^{+(2)}(p,m;h,n)$ means that all
averages $B_{q\sigma}(p,h)$ in it has a interchanged indices p
and h.

\begin{table}[h]
\caption{\label{tableR1}$R_{q\sigma}^{(1)}(p,m;h,n)$}
\begin{indented}
\item[]\begin{tabular}{@{}l|l|lll}
\br
 & & m=1 & m=2 & m=3 \\
\mr
& h=1 & $B_{q\sigma}(1,n)$ & 0 & 0 \\
p=1 & h=2 & 0 & $B_{q\sigma}(2,n)$ & 0 \\
& h=3 & $-B_{q\sigma}(3,n)$ & 0 & 0 \\
\mr
& h=1 & $B_{q\sigma}(2,n)$ & 0 & 0 \\
p=2 & h=2 & 0 & $B_{q\sigma}(2,n)$ & 0 \\
& h=3 & 0 & $-B_{q\sigma}(3,n)$ & 0 \\
\mr
& h=1 & 0 & 0 & $-B_{q\sigma}(1,n)$ \\
p=3 & h=2 & 0 & 0 & $-B_{q\sigma}(2,n)$ \\
& h=3 & 0 & 0 & 0 \\
\br
\end{tabular}
\end{indented}
\end{table}

\begin{table}[h]
\caption{\label{tableR2}$R_{q\sigma}^{+\,(2)}(p,m;h,n)$}
\begin{indented}
\item[]\begin{tabular}{@{}l|l|lll}
\br
 & & m=1 & m=2 & m=3 \\
\mr
& h=1 & $-B_{q\sigma}(n,1)$ & 0 & 0 \\
p=1 & h=2 & $B_{q\sigma}(n,2)$ & 0 & 0 \\
& h=3 & 0 & 0 & $B_{q\sigma}(n,1)$ \\
\mr
& h=1 & 0 & $-B_{q\sigma}(n,1)$ & 0 \\
p=2 & h=2 & 0 & $-B_{q\sigma}(n,2)$ & 0 \\
& h=3 & 0 & 0 & $B_{q\sigma}(n,2)$ \\
\mr
& h=1 & $B_{q\sigma}(n,3)$ & 0 & 0 \\
p=3 & h=2 & 0 & $B_{q\sigma}(n,3)$ & 0 \\
& h=3 & 0 & 0 & 0 \\
\br
\end{tabular}
\end{indented}
\end{table}

\end{document}